\documentclass{article} 

\usepackage{opticameet3} 

\usepackage{amsmath, amsthm, amssymb}
\usepackage{adjustbox}
\usepackage{float,epsfig}
\usepackage{graphicx}
\usepackage{yfonts}
\usepackage{rotating}
\usepackage{appendix}
\usepackage{caption}
\usepackage{comment}
\usepackage{enumitem}
\usepackage{color}
\usepackage{algorithm}
\usepackage[utf8]{inputenc}
\usepackage{braket}
\usepackage{subfigure}
\usepackage{qcircuit}
\usepackage[overload]{empheq}
\usepackage{hyperref}
\usepackage{xcolor}
\usepackage[english]{babel}

\usepackage{algcompatible}
\usepackage[noend]{algpseudocode}
\usepackage{mathtools}

\usepackage{xparse}

\newcommand\authormark[1]{\textsuperscript{#1}}

\begin{document}

\def \cnot{\mathrm{CNOT}}

\title{Scalable quantum circuit simulation of a chaotic Ising chain}

\author{Sabyasachi Chakraborty,\authormark{1,*} Rohit Sarma Sarkar,\authormark{2, $\dag$} and Sonjoy Majumder\authormark{1, $\ddag$}}

\address{\authormark{1} Department of Physics, Indian Institute of Technology Kharagpur, Kharagpur 721302, India\\
\authormark{2}Department of Mathematics, Indian Institute of Technology Kharagpur, Kharagpur 721302, India}

\email{\authormark{*} sabyasachi.sch@gmail.com, $^\dag$ rohit15sarkar@yahoo.com, $^\ddag$ sonjoym@phy.iitkgp.ac.in}

\begin{abstract}

The recent advancements in out-of-time-ordered correlator (OTOC) measurements have provided a promising pathway to explore quantum chaos and information scrambling. However, despite recent advancements, their experimental realization remains challenging due to the complexity of implementing backward time evolution. Here, we present a scalable quantum circuit combined with the interferometric protocol, offering a more efficient framework for OTOC measurement. Using this method, we simulate commutator growth in integrable and chaotic regimes of a 9-qubit Ising chain. Our Trotterized circuit achieves errors below $10^{-11}$ with 4th-order Trotterization and performs well even with lower-order Trotterization approximations. We believe, this approach paves the way for studying information dynamics, highly entangled quantum systems, and complex observables efficiently.

\end{abstract}

\section{Introduction}

The realization of fault-tolerant quantum computers, in principle, will provide us with a unique framework for studying complex dynamical processes of quantum many-body systems, which are challenging for classical computers \cite{lloyd1996universal, zalka1998simulating}. One such example is the study of quantum information scrambling (QIS) \cite{hayden2007black, sekino2008fast, shenker2014black}. QIS delves into the intricate processes of local information dispersal and entanglement growth in many-body quantum systems. This phenomenon provides a pathway for understanding how isolated quantum systems thermalize and plays a fundamental role in studying chaos in quantum systems \cite{maldacena2016bound, bohrdt2017scrambling}. Hence, a deeper understanding of quantum scrambling is crucial for designing more efficient quantum algorithms and improving quantum error correction techniques \cite{choi2020quantum}. 

One of the most common approaches to quantify QIS is out-of-time-ordered correlators (OTOCs) \cite{swingle2018unscrambling}. OTOCs are formulated within the Heisenberg picture, where quantum operators evolve with time, keeping quantum states unchanged. Thus, an initially local operator acquires time dependence during unitary-time evolution. Now, to compute OTOCs at each time step, the process requires both forward and backward dynamics in time. It makes the experimental measurements of OTOCs more difficult \cite{swingle2016measuring, zhu2016measurement, yao2016interferometric, yunger2017jarzynski} because of reverse the dynamic process. Although OTOCs can be simulated on classical or quantum computers, classical simulations are limited to small systems or weakly correlated models due to computational constraints. In contrast, fault-tolerant quantum computers have the potential to enable large-scale simulations of scrambling dynamics. However, current quantum devices are constrained by noise and limited connectivity, confining simulations to systems with a restricted number of qubits and small circuit depths. Thus, simulating scrambling on near-term quantum computers holds significant challenges. Furthermore, correlation measurements are usually performed with interferometric protocols \cite{swingle2016measuring, zhu2016measurement, yao2016interferometric} or weak measurements \cite{yunger2017jarzynski}. Though, individual approaches have distinct advantages and are valuable for different scenarios, the implementation of weak measurement protocols is notably more difficult than the relatively straightforward interferometric methods.  

In this work, we implement a quantum circuit approach (discussed in detail in reference \cite{sarkar2024}) using the Suzuki-Trotter approximation \cite{trotter1959product, suzuki1976generalized, PhysRevX.11.011020} and in addition, an interferometric measurement protocol is used to calculate OTOCs in a transverse field Ising model. Our time evolution quantum circuits (composed of exponential of Pauli strings of $n$ length) utilize single-qubit rotation, Hadamard, and CNOT gates. Our circuit is further based on the result that any two Pauli-string operators, consisting of identity and $X$ gates, are permutation similar, and the associated permutationally matrices can be expressed as a product of CNOT gates, with the $n$-th qubit acting as the control qubit. As a result, the proposed circuit model for the exponential of any Pauli-string operator can be implemented on a quantum hardware with low connectivity. The key feature of this model is its scalability as quantum circuits for $(n+1)$-qubit systems can be constructed from $n$-qubit circuits by adding new quantum gates and an extra qubit. In our case, implementing the backward dynamics of the circuit is comparatively simple. It can be done by only reversing the dynamics of the single-qubit rotation gate at the $n$-th qubit when the Hamiltonian consists of Pauli-$Z$ and/or $X$ string operators. For Hamiltonians that include Pauli-$Y$ string operators, the additional $R_Z(\pi/4)$ gates must be replaced with their Hermitian conjugates, $R_Z^\dagger(\pi/4)$ to implement backward dynamics. Thus, the circuit design provides an enhanced control and scalability, making it more effective for experimental measurements of OTOCs. 

Our paper is organized as follows. Section \ref{physical_intro} introduces the definitions and physical interpretation of OTOCs. The interferometric protocol is briefly discussed in Section \ref{intf_prt}. Section \ref{cktdetls} delves into the scalable quantum circuit design for any Pauli-strings of a $n$-qubit system. In Section \ref{result}, we apply our circuit model following the interferometric protocol and demonstrate the operator spreading in the integrable and chaotic regime of the transverse field quantum Ising model with an external magnetic field. We also showcase the circuit's performance for different input states along with different orders of Trotterization with varying time steps. Finally, we summarize our findings along with future perspectives in Section \ref{conclusion}.

\section{Definition and Physical interpretation} \label{physical_intro}

In quantum scrambling, quantum information spreads across multiple degrees of freedom in a system, typically due to chaotic unitary dynamics \cite{swingle2018unscrambling}. This phenomenon can be observed through the time-dependent growth of an operator, which is characterized by the finite value of the ``commutator'',

\begin{equation}
    C(t) = \left< [\hat{W}(t), \hat{V}]^\dagger [\hat{W}(t), \hat{V}] \right>,
    \label{commutator}
\end{equation}

where $\hat{V}$  and $\hat{W}$ are arbitrary Hermitian or unitary operators of the system defined by a Hamiltonian, $\mathcal{H}$, say, under Hilbert space $H$. Here, the time dependence comes from the Heisenberg representation, in which operators evolve as $\hat{O}(t) = \hat{U}^{\dagger}{(t)} \hat{O} \hat{U}{(t)}$ under a unitary time-evolution operator $\hat{U}(t) = e^{-i \hat{H}t}$. The thermal average, represented as $\langle \cdots \rangle = \text{Tr} \{\rho \cdots \}$, which accounts for the canonical ensemble of a constant particle number under some thermal density $\rho$. The Eq. \ref{commutator} can further be expressed as,

\begin{equation} \label{commutator_otoc}
\begin{split}
    C(t) & = \left< \hat{V}^\dagger \hat{W}^{\dagger}(t) \hat{W}(t) \hat{V} \right> + \left< \hat{W}^{\dagger}(t) \hat{V}^\dagger \hat{V} \hat{W}(t) \right> - 2 \mathcal{R} \left \{\left< \hat{W}^{\dagger}(t) \hat{V}^\dagger \hat{W}(t) \hat{V} \right> \right\}, \\
    & = 2 \left ( 1 -  \mathcal{R} \left \{ \mathcal{F}(t) \right\} \right),   
\end{split}
\end{equation}
Here, we have used the unitary property of the operators, i.e.,  $\hat{W}^{\dagger}(t) \hat{W}(t) = \hat{V}^\dagger \hat{V} = \hat{I}$ and expressed

\begin{equation} \label{otoc}
    \mathcal{F}(t) = \left< \hat{W}^{\dagger}(t) \hat{V}^{\dagger} \hat{W}(t) \hat{V}\right>
\end{equation}

as time-dependent Out-of-Time-Ordered Correlator (OTOC) function. Now, this commutator in quantum mechanics can be regarded as an extension of classical chaos through the relationship $1 -  \mathcal{R} \left \{ \mathcal{F}(t) \right\} \sim e^{\lambda t}$, where $\lambda$ represents the Lyapunov exponent. This expression highlights a correspondence between the phase-space Poisson brackets in classical chaos and the quantum commutator.  

For example, consider a one-dimensional Hamiltonian $\hat{H}$ system of $n$ interacting bodies with only non-zero nearest-neighbour interaction. The decay of ``OTOC'' or the growth of the ``commutator'' is directly related to the spread of quantum information, often referred to as information scrambling \cite{hayden2007black, sekino2008fast ,shenker2014black}. This spread can be quantified in the operator space by choosing $\hat{W}$ and $\hat{V}$ as operators of the system that act locally, and initially mutually commutator, i.e., $[\hat{W}(t=0), \hat{V}(0)] = 0$. However, they become non-commutative, i.e., $[\hat{W}(t), \hat{V}(0)] \neq 0$, during time evolution, where $W(t)$ can be expressed using Baker-Campbell-Hausdorff formula as,
\begin{equation}
    e^{i\hat{H}t} \hat{W} e^{-i\hat{H}t} = \hat{W} + it[\hat{H}, \hat{W}] + \frac{(it)^2}{2!}[\hat{H}, [\hat{H}, \hat{W}]] + \dots + \frac{(it)^k}{k!}\underbrace{[\hat{H}, [\hat{H}, \dots, [\hat{H}}_{k \,\, \text{nested commutators}}, \hat{W}]]] + \dots
    \label{exp_Wt}.
\end{equation}

Initially, only the site associated with $\hat{W}(t)$ is ``active'' locally and the operator $\hat{V}$ located at a distance $\ell$ from $\hat{W}(t)$ remains unaffected at $t=0$, i.e., $[\hat{W}(t), \hat{V}]= 0$. However, as time increases, the effect of $\hat{W}(t)$ eventually enters in the range of $\hat{V}$ leading $[\hat{W}(t), \hat{V}] \ne 0$. This range grows with time, affecting more sites due to the influence of higher-order terms in the expansion of Eq. \ref{exp_Wt} and the out-of-time-ordered correlator (OTOC) becomes a useful diagnostic tool for information scrambling in the system. 
 
 However, from equation \ref{otoc} we observe that the measurement of OTOC's requires reversible process and a perfect reversible process is experimentally challenging to mitigate reverse dissipation and there is no general method to overcome this. The protocol that avoids time reversal dissipation, come with limitations and hinders with their applicability to large systems \cite{jalabert2001environment, zurek2003decoherence, levstein1998attenuation}. Despite these challenges, recent advancements in the interferometric protocol \cite{li2017measuring, garttner2017measuring, wei2018exploring, mi2021information} offer promising avenues for probing OTOCs and studying scrambling in controlled experimental settings. In the next section, we will talk about the interferometric protocol briefly which provides an idea for measuring the OTOC.

\begin{figure}[ht]
    \centering
    \includegraphics[width=0.45\linewidth]{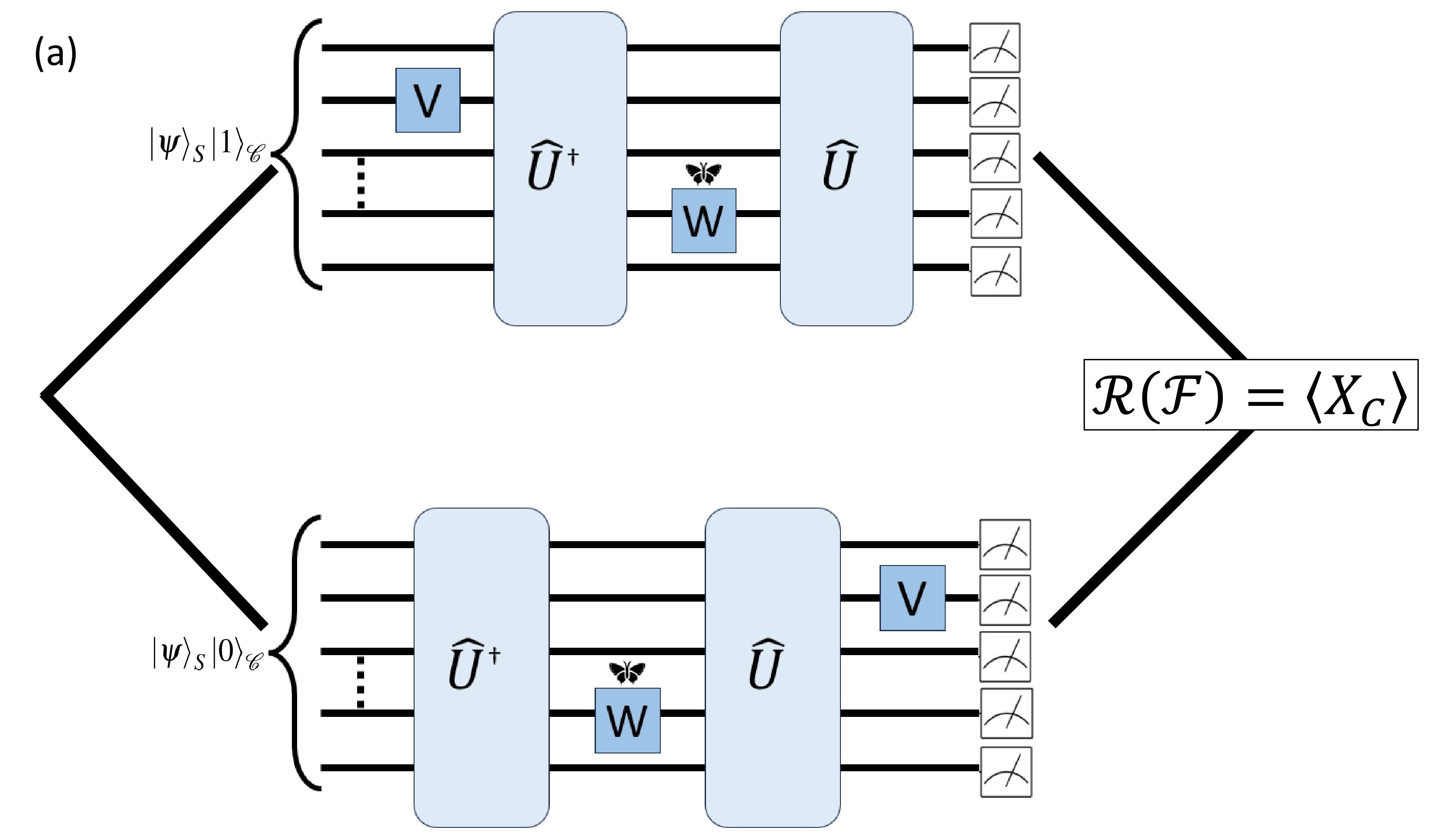}
    \includegraphics[width=0.45\linewidth]{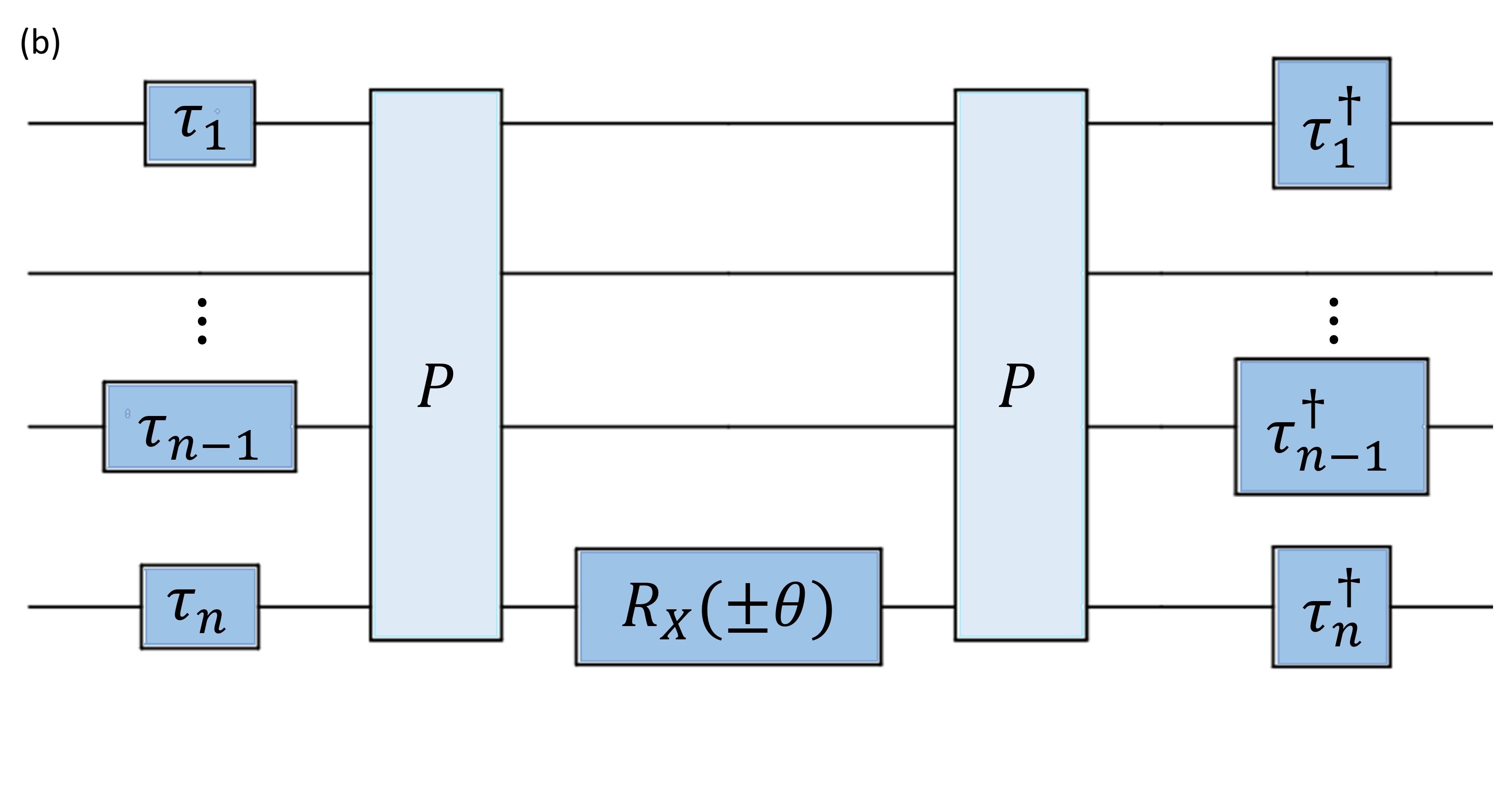}

    \caption{(a) The interferometric protocol for measuring the out-of-time-ordered correlator (OTOC). Quantum circuits $\hat{U}$ (representing forward time evolution), local operator $W$ (denoted by the butterfly symbol), and $\hat{U}^\dagger$ (representing backward time evolution) are used for time evolution of $W$ operator and applied to a quantum system consisting of qubits  $Q_1$  through  $Q_N$. A control qubit  $\mathcal{C}$, initially prepared in the state $\ket{+} = \frac{1}{\sqrt{2}} (\ket{0}+\ket{1})$, is used for arm selection of interferometer. (b) represents circuit for $\hat{U}$, implemented using a scalable quantum circuit. It is designed for $n$-qubit systems through the quantum circuit design of Pauli string at the exponent, allowing efficient simulation and measurement of OTOCs. The gates $\tau_j,j\in\{1,\hdots,n\}$ are described in subsection \ref{cktdetls} along with the permutation matrices $P$.}
    \label{ckt_protocol}
\end{figure}

\section{Interferometric protocol for measuring the OTOC} \label{intf_prt}

Here, we will briefly discuss the interferometric protocol \cite{swingle2016measuring, zhu2016measurement, yao2016interferometric}. Let $S$ represent the system of interest, which we consider a $n$-qubits spin-$\frac{1}{2}$-chain and $\ket{\psi}_s$ denote its quantum state. Let $X$, $Y$, $Z$ denote the Pauli spin operators, and the eigenstates of $\hat{Z}$ are noted as $\ket{0}$ and $\ket{1}$ with eigenvalues $+1$ and $-1$, respectively.  We know $\mathcal{F}(t)$ is a four-point function of $\hat{W}$ and $\hat{V}$ operators acting on $\ket{\psi}_s$. In the Heisenberg picture, $\hat{W}$ evolves under a time-independent Hamiltonian $\hat{H}$ as: $\hat{W}(t) = \hat{U}^{\dagger}(t) \hat{W}(0) \hat{U}(t)$, where  $\hat{U}(t)=exp(-i\hat{H}t)$.  

Now, measuring OTOC using interferometric protocol requires a control qubit $\mathcal{C}$, which is initialized in the superposition state $\ket{+}_\mathcal{C} = \frac{\ket{0}_\mathcal{C} + \ket{1}_\mathcal{C}}{\sqrt{2}}$ and the system $S$ is prepared in state $\ket{\psi}_S$. The qubit $\mathcal{C}$ controls the application of different sequences of $\hat{W}(t)$ and $\hat{V}$ operators in the two interferometer arms. If $\mathcal{C}$ contains $\ket{0}$ then $\hat{W}(t) \hat{V}$ gate-sequence will be applied on $\ket{\psi}$ state in one interferometer arm and otherwise (if $\mathcal{C}$ is occupying $\ket{1}$), $\hat{V} \hat{W}(t)$ gate-sequence will be applied in another interferometer arm. Thus, applying different order of gate sequences in the two arms of the interferometer leads to the final prepared state as

\begin{equation}
    \frac{\; \hat{U}^{\dagger}(t) \; \hat{W} \; \hat{U}(t) \; \hat{V}  \ket{\psi}_S \ket{0}_\mathcal{C} \;+ \; \hat{V} \; \hat{U}^{\dagger}(t) \; \hat{W} \; \hat{U}(t) \; \; \ket{\psi}_S \ket{1}_\mathcal{C}}{\sqrt{2}}
\end{equation}

Then we measure the control qubit in the $X$ bases, $\left< X_\mathcal{C} \right> = \mathcal{R} \left \{ \mathcal{F}(t) \right\}$, to find the real part of the OTOC.

\section{Circuit details}\label{cktdetls}

In this section, we briefly provide a quantum circuit \cite{sarkar2024} of time-evolution operator $\hat{U}=exp(\pm i\hat{H}t)$. The quantum circuit for $\hat{U}$ is obtained by developing quantum circuits for Pauli strings coupled with the Suzuki-Trotter decomposition formula where Pauli strings are defined as the Kronecker product chain of Pauli matrices and the identity matrix. To specify this, the Hamiltonian ($\hat{H}$) is written as a linear combination of local Hamiltonian ($\hat{H}_j$) as $ \hat{H}=\sum_{j=1}^{2^n}c_j \hat{H}_j$, where $H_j$'s are Pauli strings. Now, Suzuki-Trotter decomposition of the first order yields $e^{-i\hat{H}t}=\lim_{r\rightarrow\infty}(\prod_{j=1}^{2^n}e^{-ic_j \hat{H}_j\frac{t}{r}})^r$ which implies $e^{-i\hat{H}t}\approx(\prod_{j=1}^{2^n}e^{-ic_j \hat{H}_j\frac{t}{r}})^r$ for some large $r\in \mathbb{N}$. Higher-order variants of Trotterization also exist, offering improved error performance through more advanced decomposition methods and more complexity.

Hence, regardless of the Lie-product formula chosen, the quantum circuit for $\hat{U}=exp(\pm i\hat{H}t)$ requires us to construct the quantum circuit for exponential of Pauli strings. In the demonstration \cite{sarkar2024}, a generic Pauli string having $X,Y,Z$ and the Identity $I$ matrices is first converted into a Pauli string consisting of only $I$ and $X$ matrices through the use of Hadamard gates $(H)$ and rotation gates ($ S = R_Z (\frac{\pi}{4})$). This is possible due to identity relations $H^\dagger X H= Z$ and  $S^\dagger X S= Y$ (please see Theorem II.4 in reference \cite{sarkar2024}). Further, let $ \mathcal{S}^{(n)}_{I,X}$ be the set of Pauli strings containing only $X$ and $I$ matrices, then converting such strings into the block diagonal string $I^{\otimes (n-1)}\otimes X$ we multiply permutation matrices on both sides following theorem II.5 in reference \cite{sarkar2024}.

To obtain, a full mathematical description of permutation matrices, we denote two classes of permutation matrices \cite{sarkar2024} --- $\Pi \mathsf{T}^{e}_{n,x}=\prod_{j=0}^{n-2}  (\cnot_{(n,n-j-1)})^{\delta_{1,x_j} }$ and $\Pi \mathsf{T}^o_{n,x}=(\cnot_{(n-m-1,n)})(\Pi \mathsf{T}^e_{n,x})(\cnot_{(n-m-1,n)})$ , where $m$ is the greatest non-negative integer $0\leq m\leq n-2$ such that $x_{m}=1$ in the binary string of $x=(x_{n-2}\hdots x_0)$ i.e. $m=\mbox{max}\{j | \delta_{1,x_j}=1\}$ and $\delta$ denotes Kronecker delta function and $(\cnot_{(n,n-j-1)})^{0}$ is considered to be the Identity matrix. For $x=0$, we consider $\Pi \mathsf{T}^e_{n,x}$ and $\Pi \mathsf{T}^o_{n,x}$ as the identity matrix i.e. absence of any $\cnot$ gates. To understand the detailed definition of permutations matrices we refer the paper \cite{sarkar2024}.

Now, establishing the permutation similarity between Pauli strings and, in turn, their exponentials as well, we gives the quantum circuit for generic Pauli string exponentials. That is for a given any $n$-length Pauli string such that $\sigma=(\sigma_{1}\otimes \sigma_2\hdots\otimes \sigma_n)\neq I$, $\sigma$ can be written in the form  $(\sigma_{1}\otimes \sigma_{2}\hdots\otimes \sigma_n)=(\tau_{1}\otimes \tau_{2}\hdots\otimes \tau_n)^\dagger(\mu_{1}\otimes \mu_{2}\hdots\otimes \mu_n)(\tau_{1}\otimes \tau_{2}\hdots\otimes \tau_n)$ where $\tau_j$ is $H$ gate when $\mu_j$ is $X$ and $S$ gate when $\mu_j$ is $Y$ otherwise Identity gate. Then the circuit for $\exp{(\pm \iota \theta \sigma)}$ is given in Figure \ref{ckt_protocol}.b where $P$ is a permutation matrix such that $P(\mu_{1}\otimes \mu_{2}\hdots\otimes \mu_n)P=I_2^{\otimes (n-1)}\otimes X$ \cite{sarkar2024}. We note $\sigma = \tau^{\dagger} \mu \tau$  which implies $\exp{(\pm \iota \theta \sigma)} = \tau^{\dagger} \exp{(\pm \iota \mu)} \tau$. Since, $\tau = \bigotimes_{j=1}^{n} \tau_j$ is unitary and $\mu = \bigotimes_{j=1}^{n} \mu_j$.

\section{Result}\label{result}

\begin{figure}[ht]
    \centering
    \includegraphics[width=\linewidth]{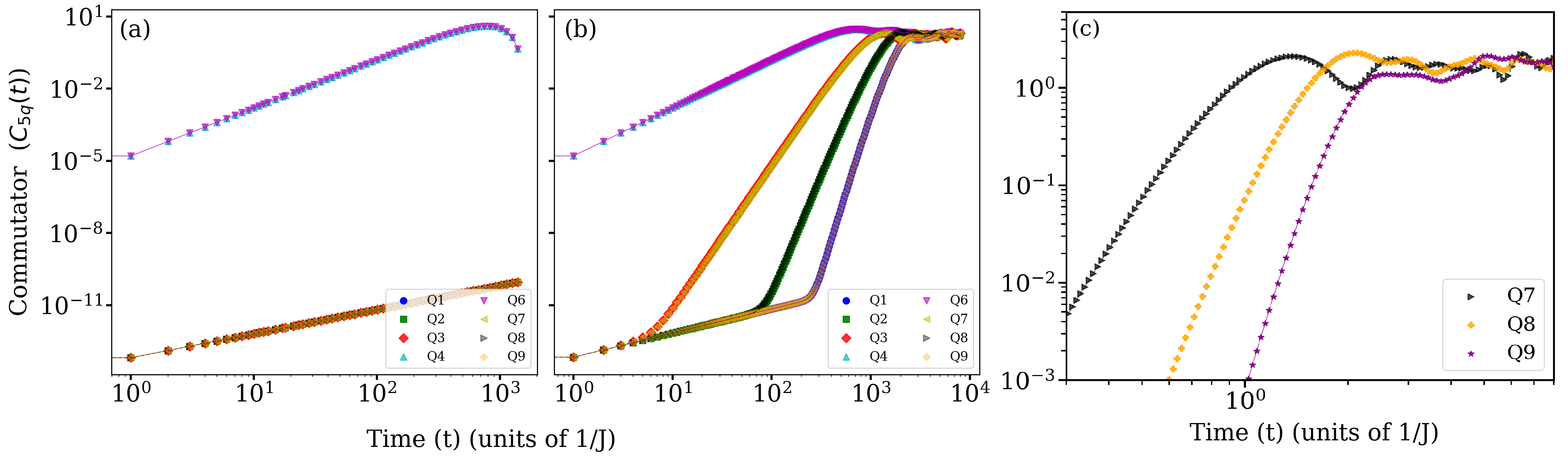}
    \caption{Commutator \( C_{5j}(t) \) versus qubit position \( j \) and time \( t \). (a) Integrable regime, taking $J=-1$, $h^Z = 1$ and $h^X = 0$, where no spreading occurs beyond nearest neighbour (Q4 and Q6). The OTOC is calculated with a minimum time step of 0.001 in units of \( 1/J \). (b) Chaotic regime, $J=-1$, $h^Z = 1$ and $h^X = 1$, showing ballistic operator spreading. (c) The growth of the commutator stops at an epoch, known as scrambling time, and exhibits small oscillations around a mean constant value. We observe at time $\sim 1$ in unit of $1/J$, the initial local information is fully scrambled. All the simulations have been carried out on a 9-qubit system using Qiskit simulator and a 4th-order Trotterization method.}
    \label{int_ch}
\end{figure}

This section uses the quantum circuit simulation (discussed in Section \ref{cktdetls}) of a one-dimensional chain of $n$-qubits Ising Hamiltonian with a transverse field under open boundary conditions. The transverse field Ising Hamiltonian for this system is described as:

\begin{equation}
    H = J \sum_{i=1}^{n-1} Z_i Z_{i+1} + \sum_{i=1}^{n} h_i^Z Z_i + \sum_{i=1}^{n} h_i^X X_i
    \label{otoc_hamil}
\end{equation}
Here, $J$ represents the interaction energy scale, $h^X$ corresponds to the transverse field strength, and $h^Z$ is the longitudinal field strength applied to the system. When $h^X$ or $h^Z$ is zero, the Hamiltonian is in an integrable regime and can be solved exactly. Otherwise, when both $h^X$ and $h^Z$ are non-zero, the system is in a non-integrable regime and shows chaotic behaviour. Specific parameter values are chosen: $n = 14$, $J = -1$, $h_i^Z = h^Z = 1$, and $h_i^X = h^X = 0$ or $=1$ depending on integrable or chaotic regime for all $i \in \{1, \cdots, n\}$.  

\begin{figure}[ht]
    \centering
    \includegraphics[width=0.6\linewidth]{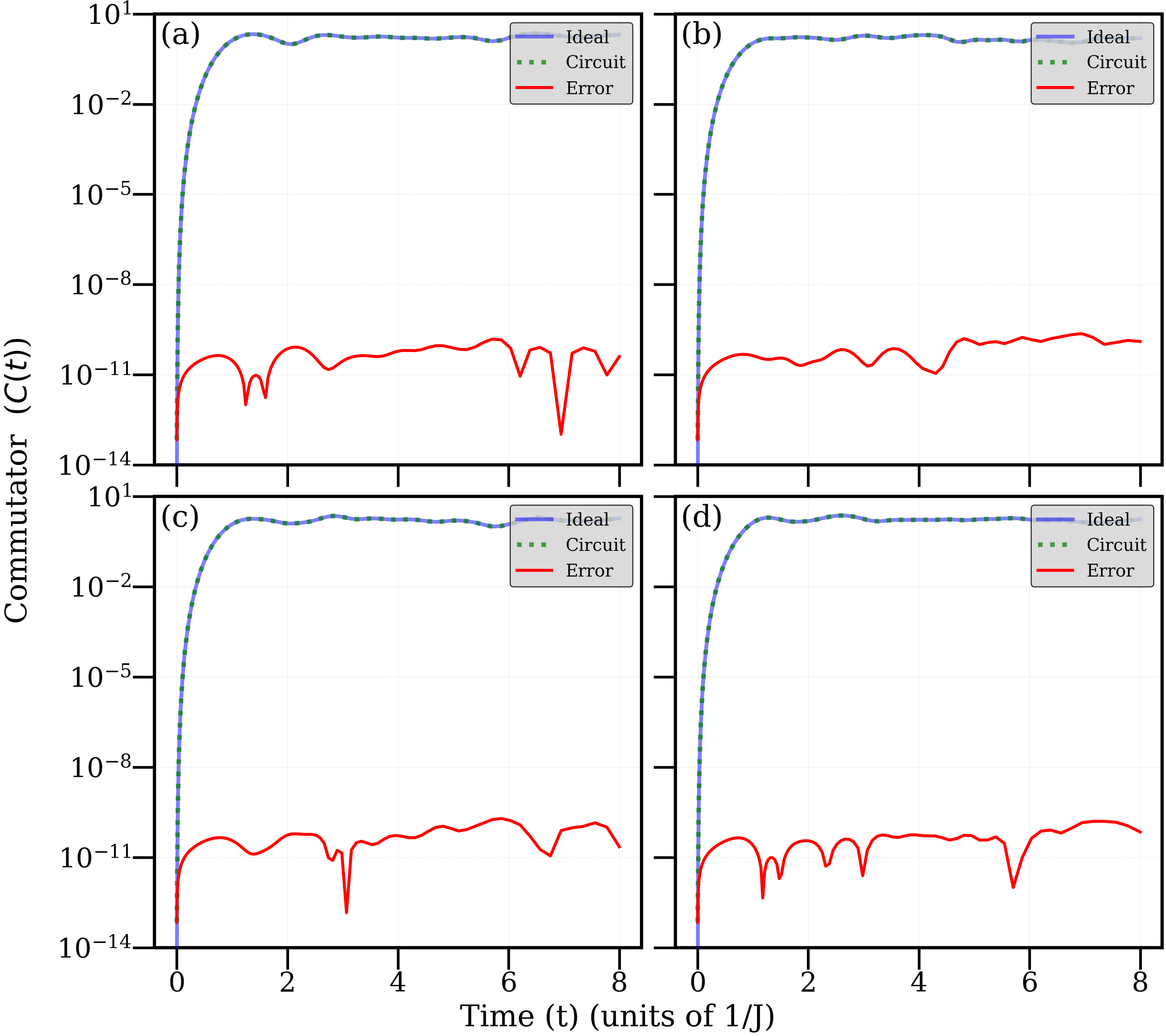}
    \caption{Comparison of commutator growth \( C_{5j}(t) \) for \( j = 3 \) across various initial states using the scalable circuit simulation with 4th-order Trotterization and direct numerical simulation with no Suzuki-Trotter approach. The initial states considered include (a) Separable state, (b) Ground state of the integrable Hamiltonian, (c) Maximally entangled GHZ state, and (d) Gaussian random distribution of \( |\pm y\rangle \) states (eigenstates of the Pauli-\( Y \) operator). The errors shown represent the norm distance between the state obtained after circuit simulation and the state obtained after direct numerical simulation, highlighting the circuit's performance under different initial conditions.}
    \label{ip_state_otoc}
\end{figure}

\subsection{Operator spreading in Integrable and Chaotic regime}
The above-described model enables efficient simulation of the Hamiltonian through Trotterization techniques. To study operator spreading, the commutator $C_{ij}(t)$ (equation \ref{commutator_otoc}) is calculated using OTOC,
\begin{equation}
    \mathcal{F_{\text{ij}}}(t) = \bra{\psi} U^\dagger X^\dagger_i U X^\dagger_j U^\dagger X_i U X_j \ket{\psi},
\end{equation}
where $\ket{\psi}$ represents the initial state, and the operator $W(t) = X(t)$ acts initially  at position $i=5$. The commutator $C_{ij}(t)$ is analyzed as a function of the position $j \in \{1, 2, \ldots, n\}$ and time $t$. $U$ represents the quantum circuit (Figure \ref{ckt_protocol}.b) simulating the unitary evolution $e^{-iHt}$. The circuit simulation protocol following the interferometric scheme is described in Figure \ref{ckt_protocol}.a. The specific choice of $W$ and $V$ in the commutator is done for the smooth dynamics and clear visualization of OTOC, making it suitable for understanding operator spreading. OTOC measurements are conducted on a 1D chain of $n = 9$ qubits. The interferometric protocol for measuring the out-of-time-ordered correlator (OTOC) involves applying a forward quantum circuit $\hat{U}$ to a system of 9 qubits, $Q_1$ through $Q_n$, followed by its inverse $\hat{U}^\dagger$ to reverse the time-evolution. A local operator $W$, represented by a butterfly symbol (in Figure \ref{ckt_protocol}.a), is introduced between the forward and backward operations to perturb the system. A control qubit \( \mathcal{C} \), initialized in the superposition state \( |+\rangle = \frac{1}{\sqrt{2}} (|0\rangle + |1\rangle) \), creates an interferometer, with \( |0\rangle \) and \( |1\rangle \) defining the two branches of interference. The forward evolution circuit \( \hat{U} \) is implemented as a scalable design suitable for \( n \)-qubit systems, enabling efficient simulation and analysis of OTOCs.

All quantum simulation results presented in this study are obtained using the Qiskit circuit simulator taking $n=9$ qubit system. To simulate the unitary-time evolution \( U = e^{-iHt} \), where both \( h^X \) and \( h^Z \) are non-zero in the Hamiltonian \( H \) (as defined in Eq. \ref{otoc_hamil}), we employed a 4th-order Trotterization scheme. The simulations were performed with a minimum time step of 0.001 in units of \( 1/J \), ensuring accuracy for studying the dynamics of the system. The scalable quantum circuit is implemented with the interferometric protocol for OTOC measurement with the initial state $\ket{0}^{\bigotimes9}$. Figure \ref{int_ch} demonstrates the spreading dynamics $C_{5j}(t)$ (where $j \in \{1,2, \cdots,9\} \setminus \{5\}$) showcasing the operator growth in the system, as observed in both integrable and chaotic regimes, respectively. In the short-time regime, before the scrambling time (see Figure \ref{int_ch}.c), the growth of the commutator remains steady and monotonic. The scrambling time is the epoch where the growth of the commutator stops. It can be understood as the time taken when the information of a system becomes fully dispersed or ``scrambled'', and the system reaches maximal entanglement entropy. 

From Figure \ref{int_ch}, we also see that beyond the scrambling time, the out-of-time-ordered correlator (OTOC) typically reaches a saturation point or stabilizes to a mean constant value, exhibiting small oscillations or fluctuations. This behaviour indicates that the system has undergone complete information scrambling, and the commutator no longer grows significantly.

\subsection{Circuit’s performance for different input states}

Now we measure the growth of the commutator $C_{5j}(t)$ for diverse initial states with fixed $j=3$ with 4th-order Trotterization approach using the scalable circuit simulation and the direct numerical simulations. The performance of the circuit simulation is compared with the direct numerical approach, where the direct approach is conducted under the assumption of ideal Hamiltonian evolution without any Trotterization error. Figure \ref{ip_state_otoc} reveals the performance of the circuits under differently prepared initial states--- (a) Separable state (\( |\psi\rangle = |\uparrow \uparrow \cdots \uparrow\rangle \)), (b) Ground state of the integrable Hamiltonian (\( H \)) obtained by setting \( h^X = 0 \) in Equation \ref{otoc_hamil}, (c) Maximally entangled GHZ state (\( |\text{GHZ}\rangle = \frac{1}{\sqrt{2}} (|0\rangle^{\otimes n} + |1\rangle^{\otimes n}) \) with \( n=9 \)), and (d) Gaussian random distribution of \( |\pm y\rangle \) states (eigenstates of the Pauli-\( Y \) operator). The randomized initial state described in Figure \ref{ip_state_otoc}(d) is, in general, used to approximate the physics of a maximally mixed state. The Figure also depicts the result, with errors represented by the norm distance between the state obtained from circuit simulation and that from direct numerical simulation. We observe that the circuit simulation errors are bound to the value of $10^{-11}$, which indicates that the circuit simulation is very close to the direct numerical approach.

\subsection{OTOC measurement with lower circuit depth}

\begin{figure}[ht]
    \centering
    \includegraphics[width=0.5\linewidth]{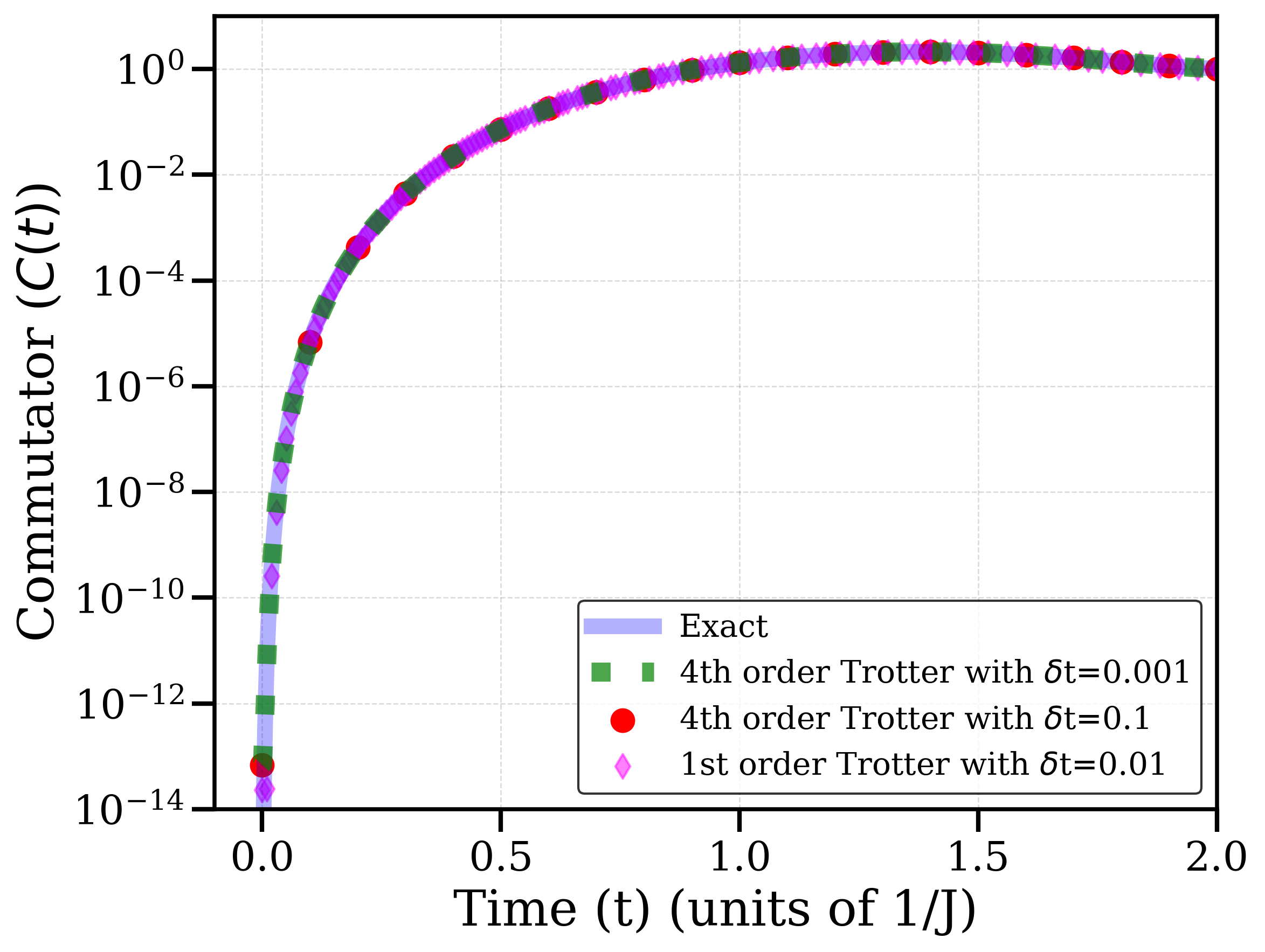}
    \caption{Plot of the commutator \(C(t)\) for a separable input state \(\ket{\psi} = \ket{\uparrow \uparrow \cdots \uparrow}\). The blue line represents the direct numerical simulation with no Suzuki-Trotter approximation. The green dotted line corresponds to the 4th-order Trotterization with \(\delta t = 0.001\). The red circles (\(o\)) denote the 4th-order Trotter with \(\delta t = 0.1\), while the magenta diamonds (\(\diamond\)) represent the 1st-order Trotterization with \(\delta t = 0.01\). Both the 4th-order Trotterization with \(\delta t = 0.1\) and the 1st-order Trotterization with \(\delta t = 0.01\) nearly align with the direct numerical simulation.}
    \label{ckt_depth_otoc}
\end{figure}

In this study, all simulations so far have used the 4th-order Trotterization with a small time step of \(\delta t = 0.001\) (in units of \(1/J\)) to ensure high accuracy. Although higher-order Suzuki-Trotter formulas reduce the approximation error, they come at the cost of increasing the quantum circuit depth. The circuit depth is defined as the number of ``layers'' of quantum gates that operate concurrently in the circuit \cite{ibmCircuitQuantum}. Therefore, there is a natural trade-off between accuracy and circuit depth.

Thus we investigate the performance of the quantum circuit simulation while reducing the time steps (\(\delta t\)) and using lower-order Trotterization schemes. For the Suzuki-Trotter product formula, the Hamiltonian in equation \ref{otoc_hamil} can be written as \(H = H_Z + H_X\), where \(H_Z = J \sum_{i=1}^{n-1} Z_i Z_{i+1} + \sum_{i=1}^{n} h_i^Z Z_i\) and \(H_X = \sum_{i=1}^{n} h_i^X X_i\). During the time evolution, while calculating the $e^{-iH \delta t}$, the Suzuki-Trotter error scale as \(\mathcal{O}(\delta t^2)\) for the 1st-order Trotterization and \(\mathcal{O}(\delta t^5)\) for the 4th-order Trotterization, where $\delta t$ represents the minimum time step. 

In Figure \ref{ckt_depth_otoc}, we plot the commutator \(C(t)\) for a separable input state \(\ket{\psi} = \ket{\uparrow \uparrow \cdots \uparrow}\). The results indicate that both the 4th-order Trotterization with \(\delta t = 0.1\) and the 1st-order Trotterization with \(\delta t = 0.01\) closely follow the exact results and significantly reduce the required circuit depth while maintaining reasonable accuracy, demonstrating the utility of the lower-order Trotterization for reducing computational resources in practical implementations using our circuit formalism.

\subsection{Advantage of the proposed circuit model}
The Hamiltonian in Eq. \ref{otoc_hamil} consists only Pauli-Z and Pauli-X terms. As a result, in the quantum circuit described in Figure 1(b), all \( \tau \) gates are replaced by Hadamard gates. The OTOC measurement demands a backward time evolution as illustrated in Figure \ref{ckt_protocol} with these gates. The choice of the transverse field Ising Hamiltonian facilitates control over various scrambling mechanisms in the circuit. Since the Hadamard gate and the permutation blocks remain unchanged under complex conjugation, that is, \( H^\dagger = H \) and \( P^\dagger = P \), the backward evolution only requires controlling the \( R_X^\dagger \) gate in the \( n \)-qubit. This design makes our circuit protocol significantly more manageable and efficient for implementing OTOC measurements.

\section{Conclusion} \label{conclusion}

In this work, we implemented a scalable quantum circuit simulation to study information scrambling with interferometric protocol in a chaotic Ising chain. Using a 9-qubit quantum system, we analyzed the growth of commutators in both integrable and chaotic regimes, highlighting the operator spreading dynamics through out-of-time-ordered correlators (OTOCs). Our results showed ballistic spread of operators in the chaotic regime and demonstrated that information scrambling reaches a saturation point at the scrambling time. The proposed circuit model with 4th-order Trotterization is in good agreement with direct numerical simulations across different initial states, achieving errors below $10^{-11}$. Additionally, we demonstrated that the circuit results of the commutator growth agree well with the direct numerical value even with lower-order Trotterization schemes reducing the circuit depth significantly. This proposed methodology is robust and can be implemented to study diverse classes of Hamiltonian operators and can be extended to explore other interconnected phenomena such as thermalization, many-body localization, and operator entanglement dynamics --- topics which we will study in future. This proposed circuit approach provides a promising framework for exploring quantum chaos and quantum information dynamics in near-term quantum devices.

\section*{Acknowledgements} S. Chakraborty acknowledge the support of the Prime Minister’s Research Fellowship (\href{https://www.pmrf.in/}{PMRF}). The authors would also like to acknowledge \href{https://hpc.iitkgp.ac.in/HPCF/paramShakti}{Paramshakti Supercomputer facility} at IIT Kharagpur—a national supercomputing mission of the Government of India, for providing the necessary high-performance computational resources.

\bibliographystyle{spphys}
\bibliography{references}

\end{document}